\documentclass[aps,prl,superscriptaddress,preprint,amsmath]{revtex4}
\usepackage{graphicx,multirow,longtable,amssymb}

\begin{document}


\title{Femtonewton Forces Can Control Protein-Meditated DNA Looping}


\author{Yih-Fan Chen}
\affiliation{Department of Biomedical Engineering, University of Michigan, Ann Arbor, Michigan 48109}
\author{J. N. Milstein}
\affiliation{Department of Physics, University of Michigan, Ann Arbor, Michigan 48109}
\author{Jens-Christian Meiners}
\affiliation{Department of Physics, University of Michigan, Ann Arbor, Michigan 48109}
\affiliation{LSA Biophysics, University of Michigan, Ann Arbor, Michigan 48109}


\date{\today}

\begin{abstract}
We show that minuscule entropic forces, on the order of 100 fN, can prevent the formation of DNA loops--a ubiquitous means of regulating the expression of genes.  We observe a tenfold decrease in the rate of LacI-mediated DNA loop formation when a tension of 200 fN is applied to the substrate DNA, biasing the thermal fluctuations that drive loop formation and breakdown events.  Conversely, once looped, the DNA-protein complex is insensitive to applied force. Our measurements are in excellent agreement with a simple polymer model of loop formation in DNA, and show that an anti-parallel topology is the preferred LacI-DNA loop conformation for a generic loop-forming construct. 
\end{abstract}


\maketitle


Since Jacob and Monod's groundbreaking work on gene regulation, the {\it lac} operon has
become a canonical example of how prokaryotic cells regulate the expression of genes in
response to changes in environmental conditions \cite{lac}.  The {\it lac} operon is responsible for the efficient metabolism of lactose in {\it Escherichia coli} bacteria ensuring that enzymes capable of digesting lactose are produced only when needed.   The {\it lac} repressor-mediated DNA loop, which is formed when tetrameric {\it lac} repressor protein binds to two {\it lac} operator sites
simultaneously, is an important part of this gene regulatory network and is crucial for the
repression of {\it lac} genes\cite{oehler}. Long range genetic regulation by DNA looping, however, is not unique to the {\it lac} operon, but appears in a variety of contexts within prokaryotes, such as the {\it ara} or {\it gal} operons and is ubiquitous within eukaryotes \cite{matthews}. While the biochemistry of these processes is generally well understood, the mechanics of the assembly and breakdown of protein-mediated DNA loops has only recently garnered much attention \cite{finzi}. In this paper, we investigate the role that tension in the substrate DNA plays in the formation and breakdown of protein-mediated DNA loops, and conclude that loop formation is acutely sensitive to entropic forces on the hundred-femtonewton scale.

Protein-mediated DNA loop formation is driven by thermal fluctuations in the DNA which bring distant operators close enough for loop closure by a protein. However, it is quite surprising that the magnitude of these fluctuations, which one can estimate as $k_BT/l_p\approx80$ fN  where $l_p$ = 50 nm is the persistence length, is much smaller than the typical piconewton forces that arise in the intracellular environment, from molecular motors or DNA-cytoskeletal attachments for example. This observation has led to predictions that forces as small as a few hundred femtonewtons are sufficient to reduce the loop formation rate by more than two orders of magnitude \cite{marko,blumbergt,yant}. Given that the cellular environment is thought to regularly subject DNA-protein complexes to large static or fluctuating forces, the cell must either use mechanical pathways to regulate genetic function, or compensate for the effects of tension to ensure the stable control of gene expression. 

To experimentally study the effects of tension on the kinetics of DNA looping, we used optical tweezers in conjunction with tethered-particle motion (TPM) measurements to investigate the formation and breakdown of LacI-mediated DNA loops under a constant stretching force. We report three main results: First, the rate of loop formation is extremely sensitive to applied tension resulting in a tenfold decrease in loop formation when increasing the tension from 60 to 183 fN. Second, the lifetime of the looped state appears to be completely unaffected by forces as large as 183 fN. Third, our measurements strongly suggest that the anti-parallel conformation is the dominant topology of a generic LacI-mediated DNA loop \cite{towles}.
\begin{figure}
\includegraphics[scale=1]{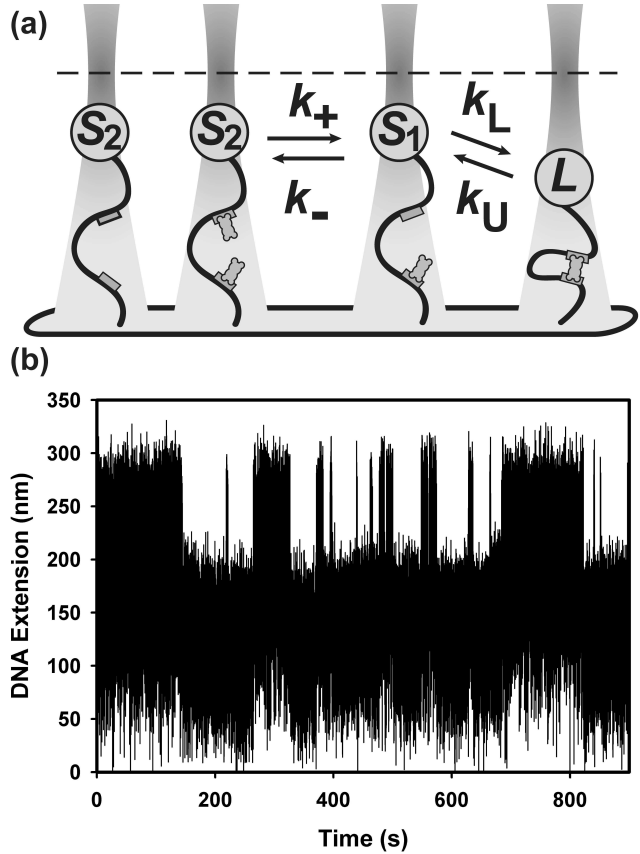}
\caption{\label{fig1}(a) Tethered DNA is trapped in the linear region of an optical potential (dashed line indicates laser focus). The figure also represents a kinetic model of DNA looping. (b) Raw experimental recording of LacI-mediated DNA looping.  The looped/unlooped threshold is chosen at the minimum between the two state distributions displayed in a histogram of the binned data.}
\end{figure}
We mechanically attenuate the thermal fluctuations that drive loop formation and breakdown, and measure the associated changes to the looping and unlooping rates, by employing a variety of optical tweezers that differs from the more conventional tweezers setup. Constant-force axial optical tweezers stretch the molecule away from the surface and trap the attached microsphere slightly below the laser focus in an approximately linear region of the optical potential. This provides effectively a constant force in the axial direction that does not change when the protein binds to or dissociates from the DNA (see Fig. 1a). Details of this set-up are described in Ref.\cite{chen1}. Because of this novel optical tweezers setup, we have been able to measure the formation and breakdown rates of LacI-mediated loops as a function of applied tensile force in the femtonewton range.
\\ 
\indent The DNA samples used in this study were prepared in a similar way to that of other TPM experiments \cite{chen1}. We surface-tethered a 1316-bp ds-DNA molecule with two symmetric {\it lac} operators spaced 305 bp apart and then attached an 800 nm polystyrene microsphere to the other end, which was trapped within the linear regime of the optical potential.  The total tension in the DNA was carefully calibrated to account for the applied optical force \cite{chen1} and volume exclusion effects arising from entropic interactions between the microsphere and the coverslip \cite{chen2}. The looping and unlooping lifetimes were measured under four different forces: 60 ($\pm$5), 78 ($\pm$6), 121 ($\pm$9), and 183 ($\pm$15) fN in the presence of 100 pM of LacI protein. In each measurement, the surface-tethered ds-DNA molecule was stretched by a contant force while the CCD camera captured defocused images of the tethered microsphere at a frame rate of 100 fps.  The looped and unlooped states of the DNA molecule, which correspond to different axial positions of the microsphere, can be measured by analyzing the resulting images, as shown in Fig. 1b.  By directly observing changes in the axial position of the microsphere, the temporal resolution for detecting loop formation and breakdown events in our experiment is as low as 300 ms, an order of magnitude better than conventional TPM.  However, as we decrease the applied tension, it becomes increasingly difficult to resolve changes in the size of the microsphere, and at zero optical force, we must resort to conventional TPM.  Moreover, even in the absence of an optical force, a residual entropic force from excluded volume effects remains. For this reason, we were not able to obtain a direct measure of the force free loop formation and breakdown rates.
\begin{figure}
\includegraphics[scale=1]{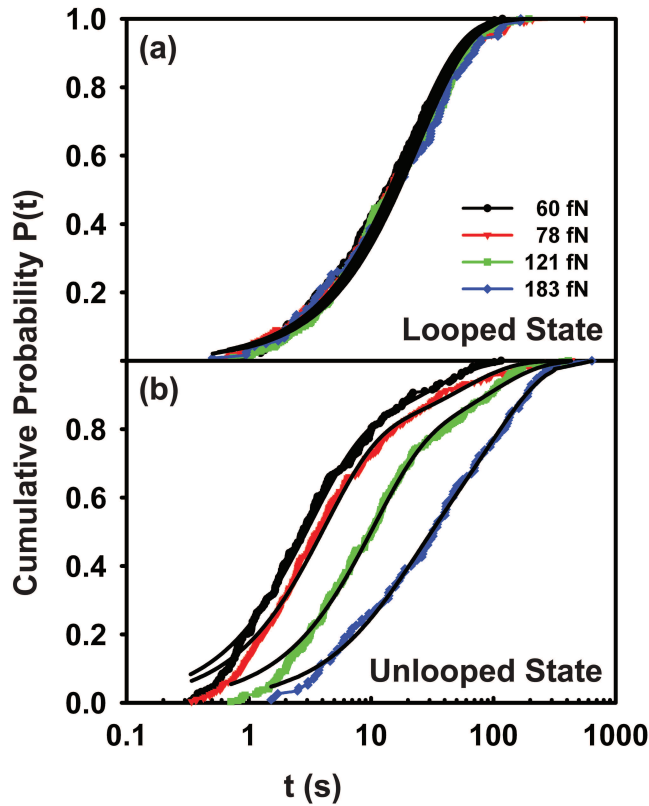}%
\caption{\label{fig2}Cumulative probability distributions of the observed durations that the DNA molecule remains in the looped and unlooped state under increasing force conditions (left to right). The data of (a) the looped state and (b) the unlooped state are fit to the single exponential function and a biexponential function respectively.  }
\end{figure}

The data was analyzed by first extracting the elapsed times between loop formation and breakdown events from time traces like the one in Fig. 1b. Then, for each force condition, the lifetime of each state was determined from a fit to the cumulative probability distribution, as shown in Fig. \ref{fig2}. The resulting distribution displayed by the looped state is well fit by a single exponential function 
\begin{equation}\label{fit1}
P(t;\tau)=1-e^{-t/\tau},
\end{equation}
with time constant $\tau$.  However, the data of the unlooped state is poorly fit to a single exponential function, but is well fit to a biexponential function
\begin{equation}\label{fit2}
P_2(t;\tau_1,\tau_2)=cP(t;\tau_1)+(1-c)P(t;\tau_2),
\end{equation}
with time constants $\tau_1$ and $\tau_2$, and dimensionless fitting parameter $c$. Results of the fits are shown in Table 1. 

One of the most striking features of the data in Fig. \ref{fig2} is that the dissociation time constant of DNA loops is unaffected by increasing the force from 60 to 183 fN. This result is in contrast to the force dependence of the time necessary to form a loop, which increases significantly with only a modest increase in applied tension. To interpret these observations quantitatively, we begin by applying a kinetic model for the underlying processes of protein binding, unbinding, loop formation, and breakdown, as illustrated in Fig. 1a.  This is the simplest model of the kinetics that is both consistent with our data and what is currently known about LacI mediated looping.

\begin{figure}
\includegraphics[scale=1]{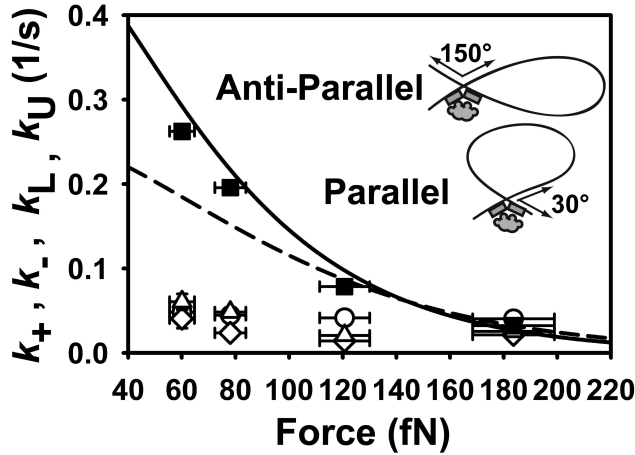} 
\caption{\label{fig3}Measured values for $k_- (\triangle)$, $k_+ (\diamond)$, $k_L (\blacksquare)$, and $k_U (\circ)$ (see Table 2). The looping rate $k_L$ is fit by the theoretical predictions for the anti-parallel (solid line) and parallel (dashed line) topologies illustrated in the insert. }
\end{figure}

If we collect all time intervals that start at a loop formation event ($L$) and end at a loop breakdown event ($S_1$), then, within this ensemble, simple first-order kinetics are given by the process
\begin{equation}
L \overset{k_U}{\longrightarrow} S_1,
\end{equation}
where $L$ is the looped state, $S_1$ is the state of the DNA with only one operator bound to a protein, and $k_U$ is the unlooping rate.  Therefore, the time dependent probability of unlooping is 
\begin{equation}
S_1(t)=1-e^{-k_Ut},
\end{equation}
which corresponds to the fit function (Eq.~\ref{fit1}) for the lifetimes of the looped state.
The kinetics of loop formation, however, are more complicated because there are different unlooped sub-states that cannot be distinguished within our experiment. We start by collecting all time intervals that begin at an unlooping event and end with the formation of a loop. The kinetics may be represented as 
\begin{equation}\label{rate}
S_2\underset{k_-}{\overset{k_+}{\rightleftharpoons}} S_1 \overset{k_L}{\longrightarrow} L,
\end{equation}
where $S_1$ is the state of one vacant and one occupied operator and may directly convert to the looped state $L$ at a rate $k_L$, or remain unlooped and convert to state $S_2$ at a rate $k_-$. State $S_2$, however, is an alternate configuration with both or neither operator occupied by a protein, which is not able to directly form a loop, but may convert to state $S_1$ at a rate $k_+$. With the initial condition $S_1(0)=1$, the first-order kinetics above may be solved for the time-dependent probability of forming a loop
\begin{equation}\label{loopP}
L(t)=1-\frac{1}{2\alpha}\left[(\kappa-k_L+\alpha)e^{-t/\tau_1}-(\kappa-k_L-\alpha)e^{-t/\tau_2}\right],
\end{equation}
where  $\kappa=k_++k_-$, $\alpha=[(\kappa+k_L)^2-4k_+k_L]^{1/2}$, and the time constants are defined as  $\tau_1=2/(\kappa+k_L-\alpha)$ and $\tau_2=2/(\kappa+k_L+\alpha)$. Equation \ref{loopP} is again a biexponential distribution and corresponds to the fit function of Eq.~\ref{fit2}.  Therefore, we can unambiguously extract the four rate constants in our kinetic model. The results are shown in Table 2 and plotted in Fig. \ref{fig3}.
\begin{table} 
\caption{\label{tab1}Fits to the cumulative probability distributions. } 
\begin{ruledtabular}
\begin{tabular}{|c|c|c|c|c|}
\hline
Force (fN) & $60\pm 5$ & $78\pm 6$ & $121\pm 9$ & $183\pm 15$\\
\hline
\multicolumn{5}{|c|}{{\bf Looped}}\\\hline
$\tau (s)$ & $20.8\pm 0.3$ & $22.5\pm 0.2$ & $24.1\pm 0.3$ & $24.8\pm 0.7$\\ \hline
\multicolumn{5}{|c|}{{\bf Unlooped}}\\\hline
$\tau_1(s)$ & $3.0\pm 0.2$ & $4.0\pm 0.42$ & $9.7\pm 0.4$ & $14.5\pm 1$\\
\hline
$\tau_2(s)$ & $31\pm 8$ & $54\pm 10$ & $91\pm 10$ & $101\pm 6$\\
\hline
$c$ & $0.77\pm 0.03$ & $0.77\pm 0.02$ & $0.74\pm 0.02$ & $0.38\pm 0.03$\\
\hline
\end{tabular}
\end{ruledtabular}
\end{table}
Our main observation is that, within the uncertainties of our measurements, $k_+$, $k_-$, and $k_U$ are independent of force whereas $k_L$ is acutely force-sensitive on the hundred-femtonewton scale. This is consistent with the conventional expectation that the rate of conversion between the unlooped states $S_1$ and $S_2$ does not vary significantly as a function of the stretching force on this scale. The insensitivity to applied force of the unlooping rate $k_U$ can be explained by considering the binding energy of the LacI protein to the DNA, whose disassociation from the $lac$ binding site is necessary to break a loop. With a binding energy of $10^{-19}$ J \cite{frank} and an operator region that spans $\sim 20$ bp, the minimum force needed to remove the protein from the operator is $\sim 10$ pN, which is several orders of magnitude greater than the tension we applied. It is clear then why the looped state is relatively insensitive to mechanical tension. On the other hand, the sensitivity of the looping rate to such small forces is quite striking and potentially rich in implications. Since the characteristic force that results from thermal fluctuations of ds-DNA is approximately 80 fN, and since DNA looping is a result of thermal fluctuations, femtonewton forces can clearly impact the loop formation process. 

Quantitatively useful models of loop formation must explicitly consider the orientation of the operators along the DNA in the looped state, as the exact geometry of the loop matters significantly. Such theories were developed by Blumberg {\it et al.}\cite{blumbergt} and, independently, by Yan {\it et al.} \cite{yant}. In this paper, we use the model developed by Blumberg {\it et al.} so begin by finding the difference in the force dependent contributions to the free energy between a looped and a stretched length of DNA:
$\Delta F=F_{L}(f,\theta)-F_{S}(f)$.  The kink angle $\theta$ is defined as the angle between the tangent vectors of the DNA at the operator sites of the protein-DNA complex.
A relation for the excess contribution to the free energy as a function of kink angle, imposed on the DNA by the loop, is given by:  
\begin{equation}\label{fkink}
\hspace{-.2cm}F_{L} = \frac{{4{f^{1/2}}(1 - \cos ({\rm{\theta }}/4))}}{{1 + 12{f^{ - 3/2}}(1 - \cos ({\rm{\theta }}/4))/(1 + \cos ({\rm{\pi }} - {\rm{\theta }}))}},
\end{equation}
where the free energy is in units of $k_BT$ and the force $f$ is in units of the characteristic force for thermal fluctuations, $f_c=k_BT/l_p\approx80$ fN.

An analytic relation for the free energy of a stretched segment of DNA is given by the difference between the potential energy of a worm-like chain (WLC) and the work done by tension:  
\begin{equation}\label{stretch}
F_{S}=-\frac{l_l x^2}{4}\left(\frac{1}{(1-x)^2}+2\right),
\end{equation}
where $l_l$ is the loop length of the DNA and $x$ is the relative extension of the DNA in units of $l_p$.  We can now calculate the characteristic time necessary to form a loop under an applied force using the principle of detailed balance
\begin{equation}\label{lifetime}
\frac{\tau_f}{\tau_0}=e^{-\Delta F},
\end{equation}
where $\tau_0$ is the characteristic time at zero force.

X-ray studies of the co-crystals of LacI protein bound to short operator fragments have revealed the structure of the DNA protein complex \cite{lewis}. These results impose constraints upon, but do not fully determine the topology of the DNA loop. The preferred direction in which the DNA enters and leaves the looped complex remains unsettled, but the corresponding topologies are either anti-parallel conformations, with a kink angle of approximately $150^\circ$, or parallel conformations, with a kink angle of $30^\circ$. 

To fit the data, we calculate  $\tau_f=1/k_L$ from Eq.~\ref{lifetime} as a function of force using the WLC model to provide the relative extension $x$ in Eq.~\ref{stretch}.  Since we cannot directly measure the force free lifetime $\tau_0$, we use this as a single adjustable parameter to fit the curves. The value for $\tau_0$ is given by a least squares fit to the data.  We then generate a curve for both the anti-parallel and parallel conformations and see, from Fig. \ref{fig3}, that the anti-parallel topology is more force-sensitive than its parallel counterpart.  Our data suggest that the anti-parallel conformation is the dominant topology of a generic LacI-mediated DNA loop.
In conclusion, our results establish that very small forces, on the order of a hundred femtonewtons, can control the assembly of the regulatory protein-DNA complex necessary for expression of the {\it lac} gene. On the other hand, once formed, the looped complexes are quite stable and cannot easily be disrupted by tension in the substrate DNA, giving the system much-needed robustness. Thus, it appears more than likely that mechanical pathways can control transcription through the application of tiny forces that are generated by other intracellular processes. We hope that the development of force measurement techniques inside living cells will lead to the identification of such pathways. We also conclude that such regulatory forces would likely act on the assembly process of these complexes, but not their breakdown. 
\begin{table}
\caption{\label{tab2} Rates extracted from kinetic rate equations.}
\begin{ruledtabular}
\begin{tabular}{|c|c|c|c|c|c|}
\hline
\multicolumn{2}{|c|}{Force (fN)} & $60\pm 5$ & $78\pm 6$ & $121\pm 9$ & $183\pm 15$\\
\hline\hline
\multirow{4}{*}{{\shortstack{Kinetic \\ Rates \\ $(10^{-3}/s)$}} 
} & $k_U$ & $48.1\pm 0.6$ & $44.5\pm 0.4$ & $41.5\pm 0.6$ & $40\pm 1$\\
\cline{2-6} & $k_L$ & $262\pm 7$ & $196\pm 6$ & $79\pm 1$ & $32\pm 1$\\
\cline{2-6} & $k_-$ & $61\pm 9$ & $48\pm 5$ & $21\pm 2$ & $25\pm 4$\\
\cline{2-6} & $k_+$ & $40\pm 10$ & $24\pm 5$ & $14\pm 2$ & $21\pm 3$\\
\hline
\end{tabular}
\end{ruledtabular}
\end{table}

\acknowledgements{We thank Jason Kahn for providing us with the LacI protein.  This work was supported by the National Institutes of Health, Grant No. GM65934, and the National Science Foundation FOCUS, Grant No. 0114336.} 

\end{document}